\documentstyle[12pt,lathuile]{article}
\begin{document}
\def\pl#1#2#3{{\it Phys. Lett. }{\bf #1}(#2)#3}
\def\plb#1#2#3{{\it Phys. Lett. }{\bf B#1}(#2)#3}
\def\zp#1#2#3{{\it Z. Phys. }{\bf #1}(#2)#3}
\def\zpc#1#2#3{{\it Z. Phys. }{\bf C#1}(#2)#3}
\def\prl#1#2#3{{\it Phys. Rev. Lett. }{\bf #1}(#2)#3}
\def\rmp#1#2#3{{\it Rev. Mod. Phys. }{\bf#1}(#2)#3}
\def\prep#1#2#3{{\it Phys. Rep. }{\bf #1}(#2)#3}
\def\pr#1#2#3{{\it Phys. Rev. }{\bf #1}(#2)#3}
\def\prd#1#2#3{{\it Phys. Rev. }{\bf D#1}(#2)#3}
\def\np#1#2#3{{\it Nucl. Phys. }{\bf #1}(#2)#3}
\def\npb#1#2#3{{\it Nucl. Phys. }{\bf B#1}(#2)#3}
\def\sjnp#1#2#3{{\it Sov. J. Nucl. Phys. }{\bf #1}(#2)#3}
\def\app#1#2#3{{\it Acta Phys. Polon. }{\bf #1}(#2)#3}
\def\aop#1#2#3{{\it Ann. Phys. }{\bf #1}(#2)#3}
\def\ar#1#2#3{{\it Ann. Rev. Nucl. Part. Sc. }{\bf #1}(#2)#3}
\def\ppnp#1#2#3{   {\it Prog. Part. Nucl. Phys. }{\bf #1} (#2) #3}
\newcommand     \ba             {\begin{eqnarray}}
\newcommand     \be             {\begin{equation}}
\newcommand     \ea             {\end{eqnarray}}
\newcommand     \ee             {\end{equation}}
\newcommand     \logfr[2]       {\log\frac{#1}{#2}}
\newcommand{\GeV}{\,{\rm GeV}}
\newcommand{\NO}{\hbox{---}}
\def\Ord{{\cal O}} 

\newcommand\abs[1]{\left| #1 \right|}
\newcommand\sss{\scriptscriptstyle\rm}
\def\lsim{\stackrel{<}{{}_\sim}}
\def\gsim{\stackrel{>}{{}_\sim}}

\thispagestyle{empty}
\begin{flushright}
GeF/TH/10-01\\
hep-ph/0106300
\end{flushright}
\vskip 3.0 true cm 

\begin{center}
{\Large\bf Search for the Higgs boson: theoretical perspectives}\\
[25 pt]
{\bf  Giovanni Ridolfi\footnote{On leave from INFN, Sezione di Genova,
 Via Dodecaneso 33, I-16146 Genova, Italy.}
}\\ [10pt]
{\em Theoretical Physics Division, CERN, CH-1211 Geneva 23, Switzerland}

\vskip 2.0 true cm
{\bf Abstract} \\
\end{center}
\noindent
We present a short review of experimental and theoretical constraints
on the mass of the Standard Model Higgs boson. We briefly illustrate
the unsatisfactory aspects of the standard theory, and we present
some general considerations about possible non-standard scenarios.
\vskip 2.0 true cm
\begin{center}
{\it Talk given at Les Rencontres de la Vall\'ee d'Aoste\\
La Thuile, Italy, March 4-10, 2001}
\end{center}
\newpage

\section{Introductory remarks}
It is now an experimentally well-established fact that the
$SU(2)\times U(1)$ gauge symmetry of electroweak interactions is
spontaneously broken to $U(1)_{\rm em}$. We observe the three
Goldstone modes corresponding to the broken part of the gauge group as
the longitudinal polarization states of $W^\pm$ and $Z^0$. However, the
details of the mechanism that induces spontaneous breaking
of the gauge symmetry are still unknown. In the original formulation of the
Standard Model, spontaneous symmetry breaking is achieved 
by means of a scalar $SU(2)$ doublet
$\phi$ with unit hypercharge, whose classical potential
\be
V(\phi)=m^2\abs{\phi}^2+\lambda\abs{\phi}^4,\;\;\;\;m^2<0
\ee
has a minimum for $\abs{\phi}^2=-m^2/(2\lambda)\equiv v^2/2$, which
corresponds, at the quantum level,  to a non-invariant ground state
(assuming the validity of perturbation theory in the scalar sector).
The value of $v$ is fixed by the measurement of the
$\beta$ decay rate, $v\simeq 246$~GeV.
Only one physical degree of freedom in the Higgs sector
remains in the spectrum, a scalar $H$ with tree-level mass
$m_H^2=2\lambda v^2$.
Large values of $m_H$ correspond to strong interactions in the Higgs
sector. Indeed, the decay width of
the standard model Higgs into a pair of gauge bosons
\be 
\Gamma(H\to VV)=\frac{3}{32\pi}\frac{m_H^3}{v^2}
\ee
becomes approximately equal to $m_H$ for $m_H\sim 1.4$~TeV. In this case, 
the Higgs boson can no longer be considered as a particle.
In the following, we will concentrate on
the conventional scenario in which the Higgs sector is within
the perturbative regime, and the Higgs boson mass is not too large.

\begin{figure}[t]
\vspace{9.0cm}
\includegraphics{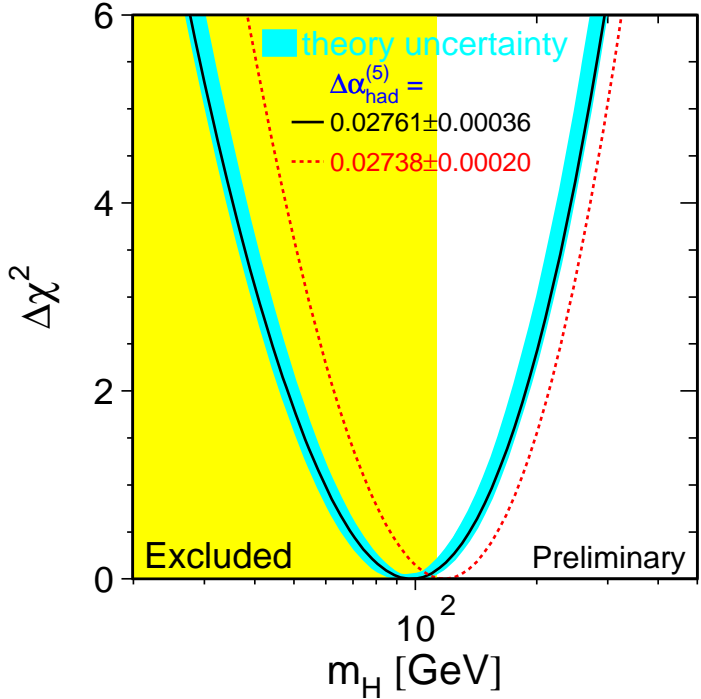}
\caption{\it Fit of the Higgs boson mass from electroweak precision data.
\label{fig:higgsfit}}
\end{figure}

The Higgs boson has not been detected so far.
The lower limit on the Higgs boson mass from direct searches is about
$113$~GeV\cite{EWWG}, but there are strong indications that, if it exists,
its mass should not be too much larger than this. In fact, a global fit to
precision observables indicates that the Higgs boson of the minimal
standard model is a light particle: the minimum value of $\chi^2$,
shown in fig.~\ref{fig:higgsfit}, is obtained for $m_H=98$~GeV, with
$m_H<212$~GeV at 95\% confidence level.

In the following, we will present a short and simple review of our present
theoretical knowledge on the Higgs boson. We will then recall the problems
that the introduction of scalar particles induces, and we will present some
considerations about how extensions of the Standard Model may appear
in future experiments.

\section{Theoretical constraints on the Higgs boson mass}
In this section, we will briefly review the arguments
that lead to conclude that values of the Standard Model
Higgs mass in the 100--200~GeV
range are favoured on  theoretical grunds.

\begin{figure}[t]
\vspace{09.cm}
\includegraphics{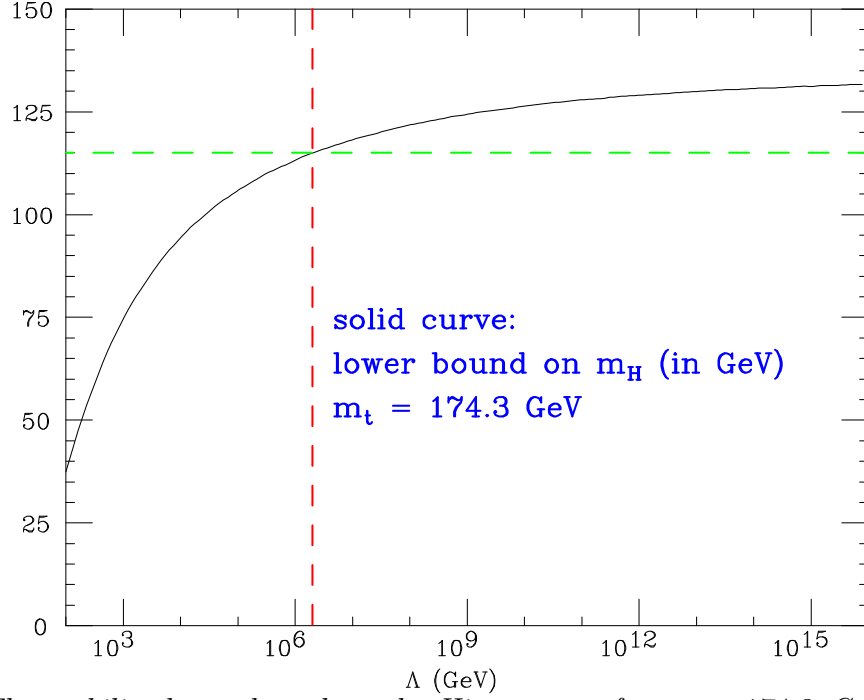}
\caption{\it The stability lower bound on the Higss mass, for $m_t=174.3$~GeV,
as a function of the cut-off $\Lambda$.}
\label{fig:lb}
\end{figure}

A lower bound on $m_H$ originates from the requirement that
the scalar potential be bounded from below even after the inclusion of
radiative corrections. In practice, it turns out that
this requirement is fulfilled if the running quartic coupling
$\lambda(\mu)$ stays positive, at least up to a certain scale $\mu\sim\Lambda$,
the maximum energy scale at which the theory can be considered reliable.
\begin{figure}[t]
\vspace{09.cm}
\includegraphics{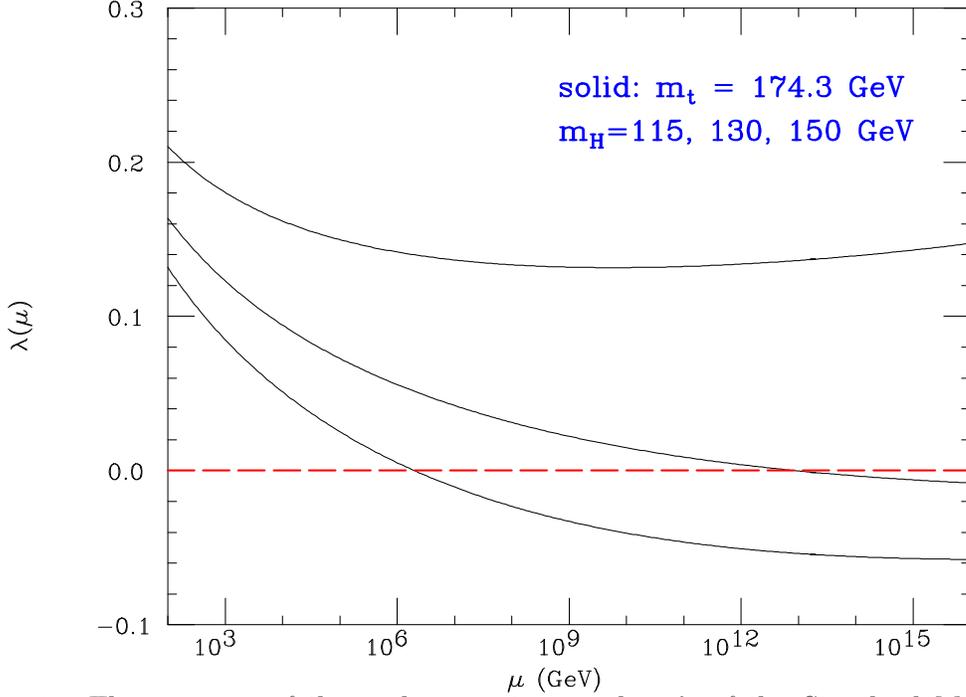}
\caption{\it The running of the scalar quartic coupling $\lambda$ of
the Standard Model, for three different values of the Higgs boson mass.}
\label{fig:run-t}
\end{figure}
In fig.~\ref{fig:run-t} the running of $\lambda$ is shown for three
different values of the initial condition, given at $\mu=m_Z$.
Clearly, the smaller is $\lambda(m_Z)$, the smaller becomes the
scale $\Lambda$ at which $\lambda$ becomes negative and the scalar potential
unbounded. Since $m_H^2\simeq 2\lambda(m_Z)v^2$, this implies 
a $\Lambda$-dependent lower bound on $m_H$, which is shown in fig.~\ref{fig:lb}
as a function of $\Lambda$ for $m_t=174.3$~GeV.
Observe that the lower bound never becomes larger than about $130$~GeV.

\begin{figure}[t]
\vspace{09.cm}
\includegraphics{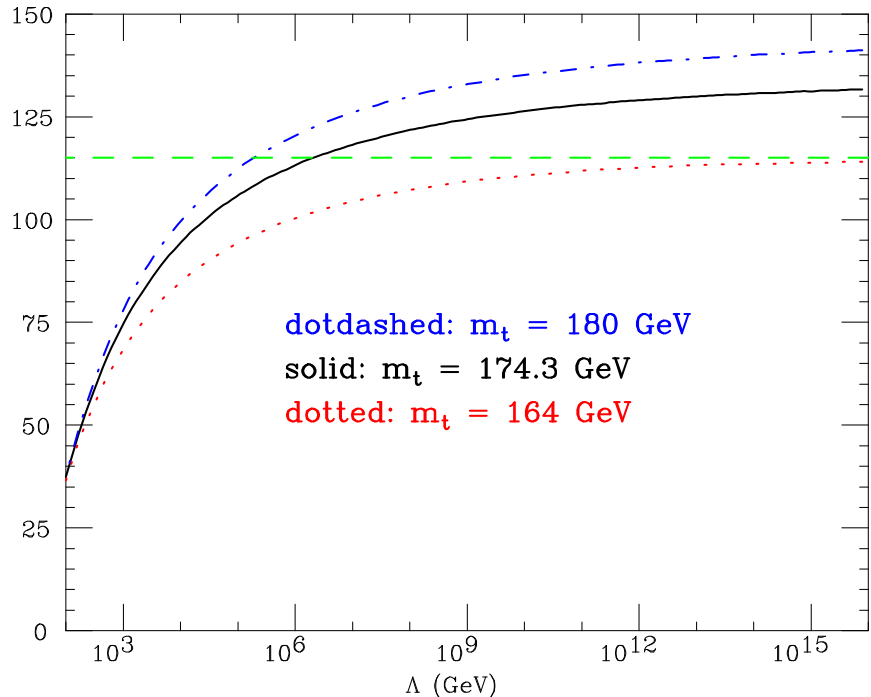}
\caption
{\it Lower bound on the Higgs mass (in {\rm GeV})
as a function of the cut-off $\Lambda$, for different values of $m_t$.}
\label{fig:lb2}
\end{figure}

As the dashed lines in fig.~\ref{fig:lb} show, a value of the Higgs
mass just above the present exclusion limit (say $m_H=115$~GeV) seems
to imply the presence of new phenomena at a scale $\Lambda\sim
10^6$~GeV. This is in fact not strictly true, for different reasons.
First, the stability lower bound of fig.~\ref{fig:lb} turns out to be
extremely sensitive to the value of the top quark mass $m_t$, which
enters the calculation because the evolution of $\lambda$ also depends
on the top Yukawa coupling. This can be seen from fig.~\ref{fig:lb2},
where the stability bound is shown for different values of $m_t$.
For $m_t=164$~GeV, which is only about two standard deviations smaller
than the central value of the Tevatron measurements, the scalar
potential of the Standard Model is bounded from below up to energies of the
order of the unification scale, $\sim 10^{16}$~GeV.

Furthermore, the stability lower bound can be released by allowing
metastability of the ground state, instead of requiring its absolute
stability, provided the lifetime of the metastable vacuum is larger
than the age of the Universe, $T\sim 10^{10}$~yrs.  The decay
probability of the false vacuum per unit volume and per unit time is
given, to one loop accuracy, by
\be
\frac{\Gamma}{V}
=\frac{e^{-S_1[h]}}{V}
=\frac{S^2_0[h]}{4\pi^2}e^{-S_0[h]}
\abs{\frac{Det'(S_0''[h])}{Det(S_0''[0])}}^{-1/2},
\ee
where $h(x)$ -- the { \it bounce} -- is the solution of classical
field equations that interpolates between the true and the metastable
vacuum state, $S_0[h]$ ($S_1[h]$) the corresponding value of the tree-level
(one-loop) euclidean action, and $Det'$ indicates that the
functional determinant is to be calculated with the zero eigenvalues omitted.
\begin{figure}[t]
\vspace{09.cm}
\includegraphics{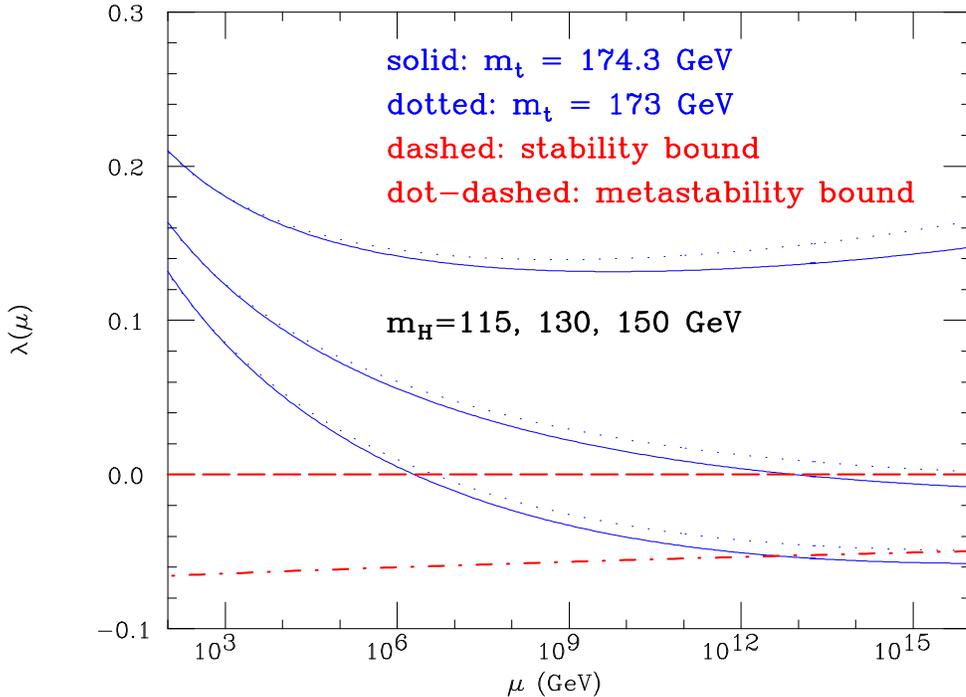}
\caption{\it Running of $\lambda$ for three different values of $m_H$
and $m_t=174.3$~GeV (solid) and $m_t=173$~GeV (dotted).
The stability (dashed) and metastability (dot-dashed) lower bounds are
also shown.}
\label{fig:run-m}
\end{figure}
The metastability bound has been computed recently to one loop
accuracy (see ref.\cite{irs} and references therein).  The results are
summarized in fig.~\ref{fig:run-m}, where the dashed and dot-dashed
curves represent the stability bound, $\lambda=0$, and the
metastability bound, respectively.  One finds that for $m_H=115$~GeV
and $m_t$ at its central value, the metastability bound is violated at
a much higher $\Lambda$ than the absolute stability bound.  If
$m_t=173$~GeV,\footnote{This value is affected by an uncertainty
of $\pm 2$~GeV, due to higher order QCD corrections.}
less than $1\;\sigma$ away from the central value,
the lower bound is respected up to the grand unification scale. Therefore,
even if the Higgs boson mass is close to its present exclusion limit,
one cannot conclude that new physics is necessarily present at
relatively low energies, on the basis of internal consistency
arguments only. 

A long-known upper bound on the Higgs mass
comes from unitarity of the scattering matrix.
Consider elastic scattering of longitudinally polarized $Z$ bosons:
\be
Z_L Z_L \to Z_L Z_L.
\ee
The corresponding amplitude, in the limit $s\gg m_Z^2$, can be computed
using the equivalence theorem:
\be
{\cal M}=-\frac{m_H^2}{v^2}\left[
\frac{s}{s-m_H^2}+\frac{t}{t-m_H^2}+\frac{u}{u-m_H^2}\right],
\ee
and the unitarity bound on the $J=0$ partial amplitude takes the form
\be
\abs{{\cal M}_0}^2\to\left[\frac{3}{16\pi}\frac{m_H^2}{v^2}\right]^2<
\frac{s}{s-4m_Z^2},
\ee
which, for $s\gg m_Z^2$, implies
\be
m_H<\sqrt{\frac{16\pi}{3}}\,v\sim 1\;{\rm TeV}.
\ee
Slightly more restrictive bounds ($\sim 800$~GeV) are obtained
considering other scattering processes, such as $Z_L W_L \to Z_L W_L$.

\begin{figure}[t]
\vspace{09.cm}
\includegraphics{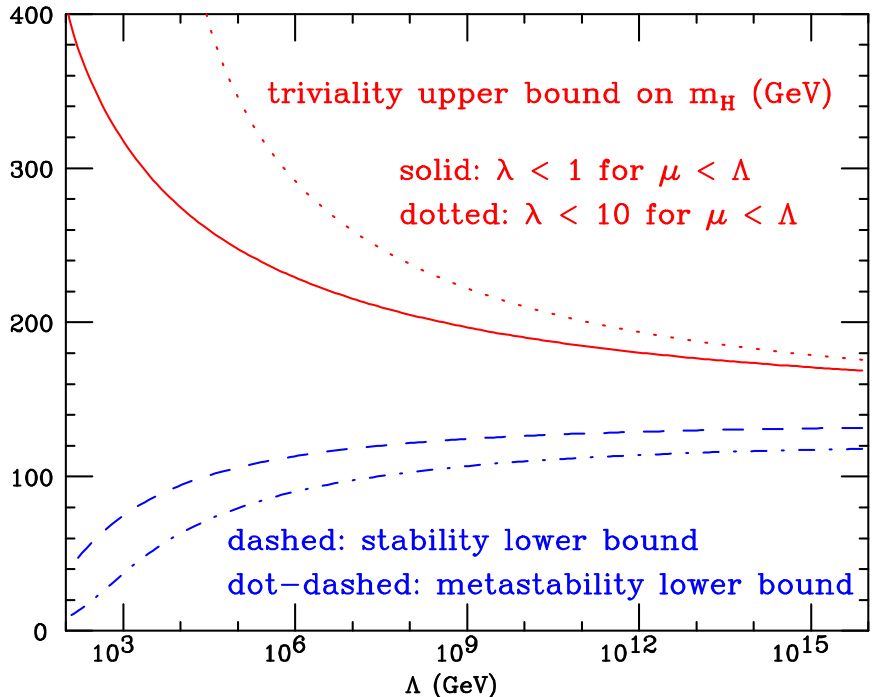}
\caption{\it Upper and lower bounds on $m_H$.}
\label{fig:bothbounds}
\end{figure}

A less rigorous, but more severe constraint is the so-called triviality
bound. The coupling $\lambda$ has a Landau pole; this can be seen explicitly
by looking at the solution of the renormalization group equation for $\lambda$
in the simplified case when gauge and Yukawa couplings are neglected. One
finds
\be
\lambda(\mu^2)=\frac{\lambda(m_Z^2)}
{1-\frac{3}{4\pi^2}\lambda(m_Z^2)\log\frac{\mu^2}{m_Z^2}}
\qquad{\rm (no\;gauge\;and\;Yukawa\;couplings)}
\ee
which has a singularity for
\be
\mu^2=m_Z^2 \exp\left[\frac{4\pi^2}{3\lambda(m_Z^2)}\right].
\ee
The location of the Landau pole in the real case is at a different
value of $\mu^2$, but the qualitative behaviour is the same.  The
theory is no longer perturbative when $\mu$ approaches this value, and
the one-loop approximation is no longer reliable. Should this
behaviour persist also at higher perturbative orders, one should
conclude that the theory is consistent at all energy
scales only if $\lambda=0$, that is, if the theory is a trivial one.
There is no rigorous proof that this is the case for the
Standard Model, but lattice computations indicate that the
$\lambda\phi^4$ theory is indeed a trivial one.  This is of course
unacceptable in the Standard Model: we need a non-zero quartic
coupling to implement the spontaneous breaking of the gauge symmetry.
Therefore, we are forced to admit that the Standard Model is only an effective
theory, valid up to some energy scale $\Lambda$, 
defined as the scale at which $\lambda(\mu)$ leaves the perturbative regime.
Larger values of the initial condition $\lambda(m_Z)$
correspond to smaller values of $\Lambda$; conversely, the requirement
that $\lambda$ stay within the perturbative domain for all scales
$\mu<\Lambda$, gives a $\Lambda$-dependent upper bound on $m_H$.
Of course, this upper bound depends on how we define the
perturbative domain, and therefore it is to some extent arbitrary.
The triviality upper bound obtained imposing the conditions
$\lambda<1$ and $\lambda<10$ are shown in fig.~\ref{fig:bothbounds}.
We see that in both cases the triviality bound is much more stringent than
the unitarity limit; an extremely severe
upper bound of about $180$~GeV is found if the validity of the Standard Model
is pushed up to the grand unification scale.

To summarize, spontaneous gauge symmetry breaking induced by the Higgs
mechanism with one scalar doublet and perturbative coupling is an
extremely appealing solution: it is relatively simple, and it is
consistent with present theoretical constraints.  Furthermore, it
should be noted that it can accommodate a consistent description (even
though not an explanation) of the observed pattern of flavor violation
(GIM suppression, FCNC phenomena, CP violation). At the same time,
there are different indications that such a theory cannot be valid up
to arbitrarily large energy scales, and that it should therefore be
considered as the low-energy approximation of some more fundamental
scenario.

\section{How will new physics look like?}
It is natural to ask what is the energy scale $\Lambda$ at which we
should expect non-standard phenomena to take place.  A related
question is whether it is possible to build a reasonable (i.e.,
consistent with data) extension of the Standard Model, where the upper
bound of about $200$~GeV on $m_H$ is evaded, and the Higgs mass is
close to the unitarity bound.  Assuming that there are no non-standard
degrees of freedom at the weak scale, one can parametrize physics at
scales well below $\Lambda$ by extending the Standard Model
lagrangian, as suggested in ref.\cite{BarbieriStrumia}, in the
following way:
\be
\label{nonsm}
{\cal L}_{\rm eff}=
{\cal L}_{\rm SM}+\sum_i \frac{c_i}{\Lambda^p}{\cal O}_i^{(4+p)},
\ee
where ${\cal O}_i^{(4+p)}$ are all the operators of dimension $4+p$,
$p\ge1$, consistent with the classical symmetries of ${\cal L}_{\rm SM}$,
that one can build with the Standard Model fields.
The upper bound on $m_H$ obtained from the global fit
to precision observables obviously holds under the assumption that
$\Lambda$ is large enough, so that the Standard Model is a good
approximation of the true theory at presently explored energies.
The non-standard term in eq.~(\ref{nonsm}) must be non-renormalizable,
since ${\cal L}_{\rm SM}$ already contains all renormalizable
operators allowed by the Standard Model symmetries.
The lowest (six) dimension operators (flavor-universal, $B,L,CP$-conserving)
are listed below:
\ba
\begin{array}{ll}
 \Ord_{WB}=(H^\dagger \tau^a H) W^a_{\mu\nu} B_{\mu\nu}
&\Ord_{H}=|H^\dagger D_\mu H|^2 \\
 \Ord_{LL}=\frac{1}{2}(\bar{L}\gamma_\mu \tau^a L)^2
&\Ord_{HL}'=i(H^\dagger D_\mu \tau^a H)(\bar{L}\gamma_\mu \tau^a L) \\
 \Ord_{HQ}'=i(H^\dagger D_\mu \tau^a H)(\bar{Q}\gamma_\mu \tau^a Q)
&\Ord_{HL} =i(H^\dagger D_\mu H)(\bar{L}\gamma_\mu L) \\
 \Ord_{HQ} =i(H^\dagger D_\mu H)(\bar{Q}\gamma_\mu Q)
&\Ord_{HE} =i(H^\dagger D_\mu H)(\bar{E}\gamma_\mu E) \\
 \Ord_{HU} =i(H^\dagger D_\mu H)(\bar{U}\gamma_\mu U)
&\Ord_{HD} =i(H^\dagger D_\mu H)(\bar{D}\gamma_\mu D).
\end{array}
\label{operators}
\ea
The authors of ref.\cite{BarbieriStrumia} have obtained the values
of $\Lambda$ for each operator in eq.~(\ref{operators}) by fitting
the data using eq.~(\ref{nonsm}) with one operator at a time, and
with fixed values of $m_H$. The results, at 95\% C.L.,
are shown in table~\ref{tabres},
where a blank means that no value of $\Lambda$ could be found such that
$\chi^2<\chi^2_{SM}+3.85$.
\begin{table*}[t]
$$\begin{array}{c|cc|cc|cc}
m_h &\multicolumn{2}{|c|}{115\GeV}&
\multicolumn{2}{|c|}{300\GeV}&
\multicolumn{2}{|c}{800\GeV}\\ 
 c_i&-1&+1&-1&+1&-1&+1\\ \hline \hline
\Ord_{WB} &9.7 & 10  & 7.5 & \NO & \NO & \NO \\
\Ord_{H}  &4.6 & 5.6 & 3.4 & \NO & 2.8 & \NO \\
\Ord_{LL} &7.9 & 6.1 & \NO & \NO & \NO & \NO \\
\Ord_{HL}'&8.4 & 8.8 & 7.5 & \NO & \NO & \NO \\
\Ord_{HQ}'&6.6 & 6.8 & \NO & \NO & \NO & \NO \\
\Ord_{HL} &7.3 & 9.2 & \NO & \NO & \NO & \NO \\
\Ord_{HQ} &5.8 & 3.4 & \NO & \NO & \NO & \NO \\
\Ord_{HE} &8.2 & 7.7 & \NO & \NO & \NO & \NO \\
\Ord_{HU} &2.4 & 3.3 & \NO & \NO & \NO & \NO \\
\Ord_{HD} &2.2 & 2.5 & \NO & \NO & \NO & \NO 
\end{array}$$
\caption{\it Fitted values of $\Lambda$ (in TeV) at 95\% C.L. for each of the
operators in eq.~(\ref{operators}), for different values of $m_H$.}
\label{tabres}
\end{table*}
Observe that the values of $\Lambda$ are generally quite large. Observe
also that fits to data are increasingly difficult with increasing $m_H$;
for most operators, no value of $\Lambda$ can be found that allows a
Higgs boson much larger than $200$~GeV.  A fit is possible, for $m_H$
in the range 300--500~GeV, for a few operators and $\Lambda$ of the order of
a few TeV. Building well-motivated models that give rise
precisely to those effective operators is a difficult task. Some
examples are known, and have been reviewed recently in
ref.\cite{PeskinWells}, in which non-standard physics compensates the
effect of a heavy Higgs, and the fit to precision data
is as good as in the Standard Model. In some cases, they
lead to observable effects at the next generation of high energy
experiments.

\section{Hierarchy, naturalness, and fine tuning}
Apart from the considerations of the previous sections, there is a
very simple reason why the Standard Model is generally believed to be
just an effective low-energy theory: at very high
energies, new phenomena take place, that are not described by the
Standard Model (gravitation is an obvious example). However, one would
like to understand why the weak scale is so much smaller than other
relevant energy scales, such as the Planck mass or the unification
scale.  This {\it hierarchy} problem is especially difficult to solve
within the Standard Model, because of the unnaturalness of the Higgs
mass.  As we have seen, we have solid arguments to believe that the
Higgs mass is of the same order of the weak scale; however, it is not
{\it naturally} small, in the sense that there is no approximate symmetry
that prevents it from receiving large radiative corrections. As a
consequence, it naturally tends to become as heavy as the heaviest
degree of freedom in the underlying theory (and therefore, maybe, of the
order of the Planck mass or of the unification scale),
unless the parameters are accurately
chosen. It is instructive to see explicitly how this phenomenon arises
in a simple example. Consider a theory of two scalars interacting
through the potential
\be V_0(\phi,\Phi)=
\frac{m^2}{2}\phi^2+\frac{M^2}{2}\Phi^2 +\frac{\lambda}{4!}\phi^4
+\frac{\sigma}{4!}\Phi^4 +\frac{\delta}{4}\phi^2\Phi^2
\ee
(which is
the most general renormalizable potential, if symmetry under
$\phi\to-\phi$, $\Phi\to-\Phi$ is required),
and assume $M^2\gg m^2>0$.  In order to check
whether this mass hierarchy is conserved at the quantum level, let us
compute one-loop radiative corrections to $m^2$ by taking the second
derivatives of the effective potential at its minimum, $\phi=\Phi=0$.
We get
\be
m^2_{\rm one\; loop}=m^2(\mu^2)
+\frac{\lambda m^2}{32\pi^2}\left(\log\frac{m^2}{\mu^2}-1\right)
+\frac{\delta M^2}{32\pi^2}\left(\log\frac{M^2}{\mu^2}-1\right),
\ee
where the running mass $m^2(\mu^2)$ obeys the renormalization group equation
\be
\label{RGE}
\mu^2\frac{\partial m^2(\mu^2)}{\partial \mu^2}=
\frac{1}{32\pi^2}\left(\lambda m^2+ \delta M^2\right).
\ee
Corrections to $m^2$ proportional to $M^2$ appear at one loop. One can choose
$\mu^2\sim M^2$ in order to get rid of them, but they reappear 
through the running of $m^2(\mu^2)$.
The only way to preserve the hierarchy $m^2\ll M^2$
is carefully choosing the parameters, so that
\be 
\lambda m^2 \sim \delta M^2,
\ee
but this requires fixing the renormalized parameters of the theory
with an unnaturally high accuracy :
\be
\frac{\delta}{\lambda} \sim \frac{m^2}{M^2}
\ee
This is what is usually called a {\it fine tuning} of the parameters.
The situation is similar when $m^2<0$, $M^2\gg\abs{m^2}>0$.
In this case, the the tree-level potential has a minimum at
$\Phi=0$, $\phi^2=-6m^2/\lambda\equiv v^2$,
and the symmetry $\phi \to-\phi$ is spontaneously broken.
The physical degrees of freedom in this case are
$\Phi$, with mass $m^2_\Phi=M^2$, and $\phi'=\phi-v$ with mass
$m^2_{\phi'}=-2m^2=\lambda v^2/3$.
At one loop, $v^2$ is given by the minimization condition
\ba
&&m^2+\frac{\lambda}{6}v^2
+\frac{1}{32\pi^2}
\left[\lambda\left(m^2+\frac{\lambda}{2} v^2\right)
\left(\log\frac{m^2+\frac{\lambda}{2} v^2}{\mu^2}-1\right)\right.
\nonumber\\
&&\phantom{m^2+\frac{\lambda}{6}v^2+\frac{1}{32\pi^2}}
\left.+\delta\left(M^2+\frac{\delta}{2} v^2\right)
\left(\log\frac{M^2+\frac{\delta}{2} v^2}{\mu^2}-1\right)
\right]=0.
\ea
Following the same procedure as in the unbroken case, one finds
\be
m^2_{\phi'}=\frac{\lambda v^2}{3}
+\frac{v^2}{32\pi^2}\left[\lambda^2 \log\frac{m^2+\frac{\lambda}{2}v^2}{\mu^2}
+\delta^2\log\frac{M^2+\frac{\delta}{2}v^2}{\mu^2}\right]
\ee
with $v\sim M$ without a suitable tuning of the parameters. These
simple examples show that squared masses of scalar particles receive
radiative corrections proportional to the squared masses of the
other degrees of freedom in the theory. Therefore, without a suitable
fine tuning of the parameters, they naturally become as large as the
largest energy scale in the theory. This is related to the fact that
no extra symmetry is recovered when scalar masses vanish, in contrast
to what happens, for example, for fermion masses.

In the case of the Standard Model Higgs, we are already faced
with this problem, as pointed out in ref.\cite{BarbieriStrumiaLep}.
The correction to $m^2_H$ due to a loop of top quarks is given by
\be
\label{mhcorr}
\delta m_H^2({\rm top})=\frac{3G_F m_t^2}{\sqrt{2}\pi^2}\Lambda^2
\simeq(0.27\,\Lambda)^2,
\ee
where we are assuming that the scale $\Lambda$ that characterizes
non-standard physics acts as a cut-off for the loop momentum.
We have seen in the previous section that, if one assumes that no new
degrees of freedom are present around the Fermi scale,
$\Lambda$ cannot be smaller than a few TeV.
With  $\Lambda\sim 5$~TeV eq.~(\ref{mhcorr}) gives
\be
\delta m_H^2({\rm top})\sim (1.5~{\rm TeV})^2,
\ee
which is two orders of magnitude larger than the indirect value
of $m_H$ from the global fit to precision observables.
There is an apparent paradox: precision tests favour a small
value for the Higgs mass, but at the same time, through the analysis with
effective operators, indicate that the scale $\Lambda$ of non standard
physics is too large to be compatible with the fitted value
of $m_H$.

Supersymmetry offers a solution to the naturalness problem,
provided the mass splittings within supermultiplets are not much
larger than the Fermi scale. In fact, in supersymmetric models
quadratically divergent radiative corrections to scalar masses 
are absent, as a consequence of the fact that supermultiplets
contain both bosonic and fermionic degrees of freedom.
In particular, the contribution to $m_H^2$
of a loop of {\it s-top} $\tilde{t}$ has the effect of replacing
$\Lambda^2$ in eq.~(\ref{mhcorr}) with
\be
m^2_{\tilde{t}}\log\frac{\Lambda^2}{m^2_{\tilde{t}}}
\ee
without affecting fits to precision observables. This is the strongest
argument in favour of supersymmetry at the weak scale. 

The Higgs sector of supersymmetric models has some specific features.
At least two Higgs doublets must be introduced; their neutral
components take non-zero vacuum expectation values $v_1,v_2$
(the notation $\tan\beta=v_2/v_1$ is usually adopted). After
spontaneous breaking of the gauge symmetry, five physical degrees of
freedom are left in the spectrum, usually denoted by $h$, $H$, $A$
(neutral) and $H^\pm$ (charged). The quartic scalar coupling $\lambda$
in supersymmetric models is replaced by a combination of the squared
weak gauge couplings $g,g'$. This has two important consequences: first,
the scalar potential is bounded from below by construction; second,
Higgs and weak vector boson masses are related. 
At tree level, it can be shown that, in a wide class
of supersymmetric models, the lightest Higgs $h$ must be lighter
than the $Z$ boson. Radiative corrections shift
this upper bound by an amount proportional to $G_F m_t^2$; for
$\tan\beta\gsim 4$ one finds \be
\label{ERZ}
m_h^2\simeq m_Z^2
+\frac{3}{\sqrt{2}\pi^2}G_F m_t^4\logfr{m^2_{\tilde{t}}}{v^2},
\ee
where $m_{\tilde{t}}$ is the mass of the scalar partners of the top
quark. For $m_{\tilde{t}}=1$~TeV, eq.~(\ref{ERZ}) gives 
$m_h\simeq 118$~GeV.

\section{Conclusions and outlook}
We have reviewed the basic features of the Standard Model Higgs boson,
and compared them with current experimental information.  The present
exclusion limit for the Higgs bosons from direct searches is
$m_H>113$~GeV, but there are many indirect indications that the mass
of the Standard Model Higgs may lie just above this limit, and most
likely below $\sim 200$~GeV. 

The possibility of a Higgs boson with a larger mass, whose effects are
compensated by some kind of non-standard physics, has also been
investigated in the literature, and reviewed here. It seems quite unlikely
that this could happen in well-motivated theories,
but the possibility is not ruled out.

A Higgs boson with a mass in the range 100--200~GeV is affected by the
problem of the hierarchy between the weak scale and the scale of new physics,
which is known to be larger than $\sim 5$~TeV if it is assumed that
no new degree of freedom is present at accessible scales.
The hierarchy problem has now become so compelling that it can be
cast in the form of a paradox. Supersymmetry at the Fermi scale is still
the most appealing candidate for its solution.  

\section*{Acknowledgements}
I wish to thank Gino Isidori, Chiara Mariotti and Alessandro Strumia for
discussions and suggestions. I thank the organizers of
{\it Les Rencontres de la Vall\'ee d'Aoste} for their kind hospitality.

\end{document}